\documentclass[preprint,aps,prc,showpacs,nofootinbib,secnumarabic,floatfix]{revtex4-1}

\usepackage{graphicx}
\usepackage{bm}
\usepackage{color}
\usepackage{gensymb}
\usepackage{amsmath}
\usepackage{slashed}
\usepackage{appendix}
\usepackage{hyperref}

\newcommand{\bvec}[1]{\ensuremath{\boldsymbol{#1}}}

\begin{document}

\title{Color transparency effects on the deuteron spin observables in the hard $pd \to ppn$ reaction}  

\author{A.B. Larionov}
\email{e-mail: larionov@theor.jinr.ru}
\affiliation{Bogoliubov Laboratory of Theoretical Physics, Joint Institute for Nuclear Research, 141980 Dubna, Russia}
                          
\begin{abstract}
  The pionless breakup of a deuteron induced by 15 GeV/$c$ beam proton is considered theoretically in the kinematics
  where both outgoing protons are fast and the neutron is slow in the rest frame of the deuteron. 
  Significant influence of color transparency on the deuteron vector and tensor analyzing powers is expected.
\end{abstract}

\maketitle

\section{Introduction}
\label{intro}

The color transparency (CT) effect is known as the reduction of the interaction with the surrounding nuclear medium for point-like color-singlet (anti)quark configurations
in the initial or final state of a hard two-particle collision.
CT has been predicted for binary semi-exclusive processes with large momentum transfer $h+A \to p+(A-1)^*$ \cite{Brodsky,Mueller}.
The reviews on CT can be found in Refs.~\cite{Frankfurt:1994hf,Jain:1995dd,Dutta:2012ii}. So far the experimental search for CT was mostly focused on the intermediate energy
region where BNL and JLab operate. CT was observed at AGS@BNL in the $^{12}{\rm C}(p,pp)$ reaction at $p_{\rm lab}=5.9\!-\!14.5$ GeV/$c$ \cite{Leksanov:2001ui}, and also at JLab
in the $A(e,e^\prime \pi^+)$ reaction at $E_{\rm beam}=4.0\!-\!5.8$ GeV \cite{Clasie:2007aa} and in the $A(e,e^\prime \rho^0)$ reaction
at $E_{\rm beam}=5$ GeV \cite{ElFassi:2012nr} for several nuclear targets.
In all existing experimental studies, the observable sensitive to the CT effect was nuclear transparency, $T=\sigma/\sigma_{\rm IA}$,
where $\sigma$ is the measured cross section of the studied process and $\sigma_{\rm IA}$ is the same cross section, but calculated in the impulse approximation (IA).
Thus, enhanced $T$ above predictions of the Glauber model points to the presence of the CT effect. The hardness of the underlying elementary reactions was reached
by setting $Q^2 > 1$ GeV$^2$. On the other hand, recent JLab data for $^{12}{\rm C}(e,e^\prime p)$ at $Q^2=8\!-\!14.2$ GeV$^2$, $E_{\rm beam}=6.4,10.6$ GeV
\cite{HallC:2020ijh} do not indicate CT. This controversial situation calls for new theoretical and experimental studies of CT. One idea is to use deuteron target,
where the uncertainties of the wave function are minimal and the processes $d(e,e'pn)$ \cite{Anisovich:1993eh,Frankfurt:1994kt,Li:2022uvf}
or $d(p,ppn)$ \cite{Frankfurt:1996uz,Larionov:2022gvn}  can be studied fully exclusively. In particular, it was shown in Refs.~\cite{Frankfurt:1994kt,Larionov:2022gvn}
that the nuclear transparency and deuteron tensor analyzing power (DTAP) are sensitive to the presence of CT. The consideration was, however, limited to the component
$A_{zz}$ with $z$-axis along the proton (in Ref.~\cite{Frankfurt:1994kt} - virtual photon) momentum. The purpose of this work is to extend the calculations
of Ref.~\cite{Larionov:2022gvn} of hard $pd \to ppn$ process to other components of the DTAP and also to study the deuteron vector analyzing power (DVAP).

The paper is structured as follows. Section \ref{model} contains a short description of theoretical model. Numerical results of calculations of the DTAP and DVAP
are given in Section \ref{results}. The summary and experimental opportunities are included in Section \ref{summary}.   

\section{The model}
\label{model}

The calculations were based on the generalized eikonal approximation (GEA) method developed in Refs.~\cite{Frankfurt:1996uz,Sargsian:2001ax,Larionov:2022gvn,Larionov:2024ynh}.
The starting point was Feynman diagrams describing one-, two-, and three-step scattering processes. In the simplest one-step, i.e. IA, diagram
the incoming proton experiences hard elastic scattering on the proton of the deuteron without any distortion from initial- and final-state interactions (ISI/FSI).
The two- and three-step diagrams include soft elastic scattering amplitudes of the incoming and outgoing protons on the spectator neutron. These diagrams were calculated
by factorizing the hard amplitude $M_{\rm hard}(s_{\rm hard},t_{\rm hard})$ out of the momentum integrals that were taken in the pole approximation, i.e. putting the intermediate spectator neutron
on the mass shell. In the present calculations, the latest model version \cite{Larionov:2024ynh} was used that includes more accurate three-step amplitudes with soft rescattering
of both outgoing protons, taking into account the renormalization due to the infinite sequence of subsequent rescattering, as well as improved $\Delta$'s, which are included
in the phases of the integrals in coordinate space in partial two- and three-step amplitudes.

The CT effects were included within the quantum diffusion model \cite{Farrar:1988me,Frankfurt:1994kt} by using position-dependent soft $pn$-scattering amplitudes
\begin{equation}
    M_{\rm el}(|\bvec{p}_j|,t,l)= 2 |\bvec{p}_j|m \, \sigma_{pn}^{\rm eff}(l) \,
     (i+\rho_{pn}) e^{B_{pn}t/2}  \,
     \frac{G(t\cdot \frac{\sigma_{pn}^{\rm eff}(l)}{\sigma_{pn}^{\rm tot}})}{G(t)}~,      \label{M_el}
\end{equation}
where $\bvec{p}_j$ is the momentum of the incoming ($j=1$) or outgoing ($j=3,4$) proton in the deuteron rest frame (r.f.),
$\rho_{p n}=\mbox{Re}M_{\rm el}(t=0)/\mbox{Im}M_{\rm el}(t=0)$
is the ratio of the real part of the forward $pn$ scattering amplitude to the imaginary one,
and $B_{\rm p n}$ is the slope of the dependence on the transverse momentum transfer,
$m$ is the nucleon mass.
\begin{equation}
   \sigma_{pn}^{\rm eff}(l)
  = \sigma_{pn}^{\rm tot}\left(\left[ \frac{l}{l_c}
    + \frac{\langle n^2 k_t^2\rangle}{Q^2} \left(1-\frac{l}{l_c}\right) \right]
    \Theta(l_c-l) +\Theta(l-l_c)\right)                                   \label{sigma_pn_eff}
\end{equation}
is the effective $pn$ cross section that depends on the distance $l =|(\bvec{r}_2-\bvec{r}_s)\bvec{p}_j|/|\bvec{p}_j|$
traveled by the incoming or outgoing proton between its hard interaction with a proton of the deuteron with position $\bvec{r}_2$
and soft rescattering on the neutron of the deuteron with position $\bvec{r}_s$.
\begin{equation}
  l_c = \frac{2p_j}{\Delta M^2}~,   \label{l_c}
\end{equation}
is the coherence length with parameter $\Delta M^2 \simeq 1$ GeV$^2$ as determined from pion transparency studies \cite{Clasie:2007aa}
and $\Delta M^2 \simeq 2\!-\!3$ GeV$^2$ as determined in Ref.~\cite{Li:2022uvf} that accommodates the proton transparency data \cite{HallC:2020ijh}.  
$Q^2=-t_{\rm hard}$ is the hard scale, $\sqrt{\langle k_t^2\rangle}=0.35$ GeV/$c$ is the average transverse momentum of a quark in a proton,
$n=3$ is the number of valence quarks in the proton.
\begin{equation}
  G(t)=\frac{1}{(1-t/0.71~\mbox{GeV}^2)^2}~,     \label{SachsFF}
\end{equation}
is the Sachs electric form factor of a proton. One sees from Eq.(\ref{sigma_pn_eff}) that at $l < l_c$ the effective cross section
is smaller than phenomenological total $pn$ cross section  $\sigma_{pn}^{\rm tot}$ due to CT, while at $l \geq l_c$ CT effects disappear
and $\sigma_{pn}^{\rm eff}(l)=\sigma_{pn}^{\rm tot}$.

\section{Results}
\label{results}

The DVAP and DTAP are defined, respectively, as (c.f. Refs.~\cite{Platonova:2010wjt,Uzikov:2020zho})
\begin{eqnarray}
  A_{\alpha} &=& \frac{\mbox{Sp}(MS_{\alpha}M^\dag)}{\mbox{Sp}(MM^\dag)}~,~~~\alpha=x,y,z~,     \label{A_alpha}\\
  A_{\alpha\beta} &=& \frac{\mbox{Sp}(MS_{\alpha\beta}M^\dag)}{\mbox{Sp}(MM^\dag)}~,~~~\alpha,\beta=x,y,z~, \label{A_alphabeta}
\end{eqnarray}
where $M$ is the invariant matrix element of the studied reaction process (in the present case $pd \to ppn$),
$S_{\alpha}$ are the deuteron spin matrices
and $S_{\alpha\beta} = \frac{3}{2}(S_\alpha S_\beta  + S_\beta S_\alpha) - 2\delta_{\alpha\beta}$ is the spin-quadrupole operator.
The DTAP satisfies the condition $\sum_\alpha A_{\alpha \alpha} = 0$.

The DVAP and diagonal DTAP can be expressed in terms of the reaction cross section
$\sigma_\alpha(\lambda)$ for the deuteron spin projection $\lambda$ on the axis $\alpha$:  
\begin{eqnarray}
  A_\alpha        &=& \frac{\sigma_\alpha(+1)-\sigma_\alpha(-1)}{\sigma_\alpha(+1)+\sigma_\alpha(-1)+\sigma_\alpha(0)}~,                   \label{A_a}\\
  A_{\alpha \alpha} &=& \frac{\sigma_\alpha(+1)+\sigma_\alpha(-1)-2\sigma_\alpha(0)}{\sigma_\alpha(+1)+\sigma_\alpha(-1)+\sigma_\alpha(0)}~. \label{A_aa}
\end{eqnarray}
Being, respectively, a pseudo-vector and a tensor, the DVAP and DTAP can be generally represented as
\begin{eqnarray}
  \bvec{A} &=& \sum_{i,j} f_{ij} [\bvec{p}_i \times \bvec{p}_j]~,   \label{Avec}\\
  A_{\alpha\beta} &=&  \sum_{i,j} g_{ij}\, (3p_{i,\alpha} p_{j,\beta} - \bvec{p}_i\bvec{p}_j\delta_{\alpha\beta})~, \label{Atens}
\end{eqnarray}
where the sums run over all pairs of particles $i,j$ involved in the reaction and the coefficients $f_{ij},g_{ij}$ depend on the scalar products of the
momenta of particles.

In the IA, for a spin-independent hard $pp$ amplitude, the DVAP disappears, while the DTAP is fully determined by the deuteron wave function:
\begin{equation}
  A_{\alpha \beta}^{\rm IA} 
            = \frac{(3 p_{s,\alpha}p_{s,\beta}/p_s^2 - \delta_{\alpha\beta})(\sqrt{2}u(p_s)w(p_s) - w^2(p_s)/2)}{u^2(p_s) +w^2(p_s)}~,      \label{A_ab^IA}
\end{equation}
where $\bvec{p}_s$ is the momentum of spectator neutron in the deuteron r.f., $u(p_s)$ and $w(p_s)$ are, respectively, the $S$- and $D$-wave components
of the deuteron wave function.

The final state of the reaction $pd \to ppn$ is characterized in the deuteron r.f. by the light cone variable $\alpha_s = 2(E_s-p_s^z)/m_d$,
the transverse momentum $p_{st}$ of the spectator neutron, the Mandelstam variable $t=(p_1-p_3)^2$, and the relative azimuthal angle
$\phi=\phi_3-\phi_s$ between the scattered proton and the spectator neutron.
Here, $E_s=\sqrt{\bvec{p}_s^2+m^2}$ and $m_d$ are, respectively, the energy of the spectator neutron and the deuteron mass.
The neutron is supposed to move in the $(x,z)$ plane, i.e. $\phi_s=0$.
The calculations were performed for a beam momentum $p_{\rm lab}=15$ GeV/$c$
at $t=(2m^2-s/2)(1-\cos\Theta_{\rm c.m.})$ with a fixed center-of-mass scattering angle $\Theta_{\rm c.m.}=53$ deg.. Here, $s=(E_1+m_d-E_s)^2-p_{st}^2-(p_{\rm lab}-p_s^z)^2$,
where $E_1=\sqrt{p_{\rm lab}^2+m^2}$ is the energy of the incoming proton.
This corresponds to $Q^2 \sim 5.2$ GeV$^2$. The transverse kinematics of the neutron was considered
by requiring $\alpha_s=1$. 

\begin{figure}[t!]
  \begin{center}
    \begin{tabular}{ccc}
       \includegraphics[scale = 0.3]{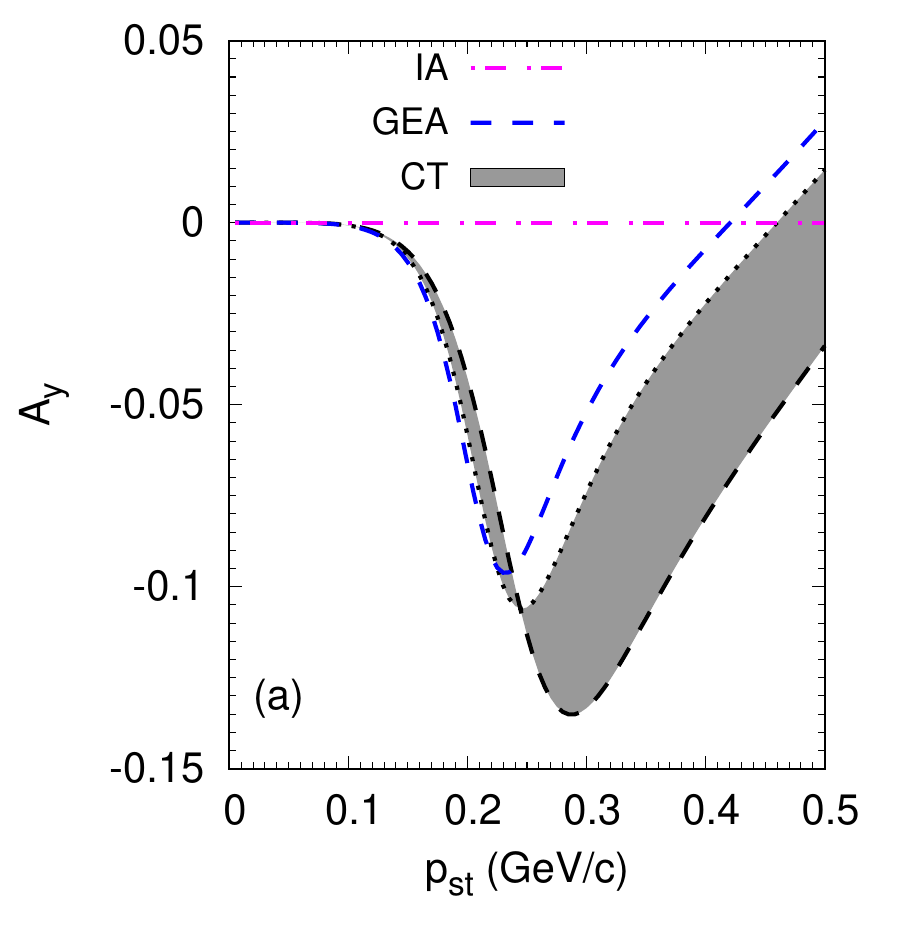} &
       \includegraphics[scale = 0.3]{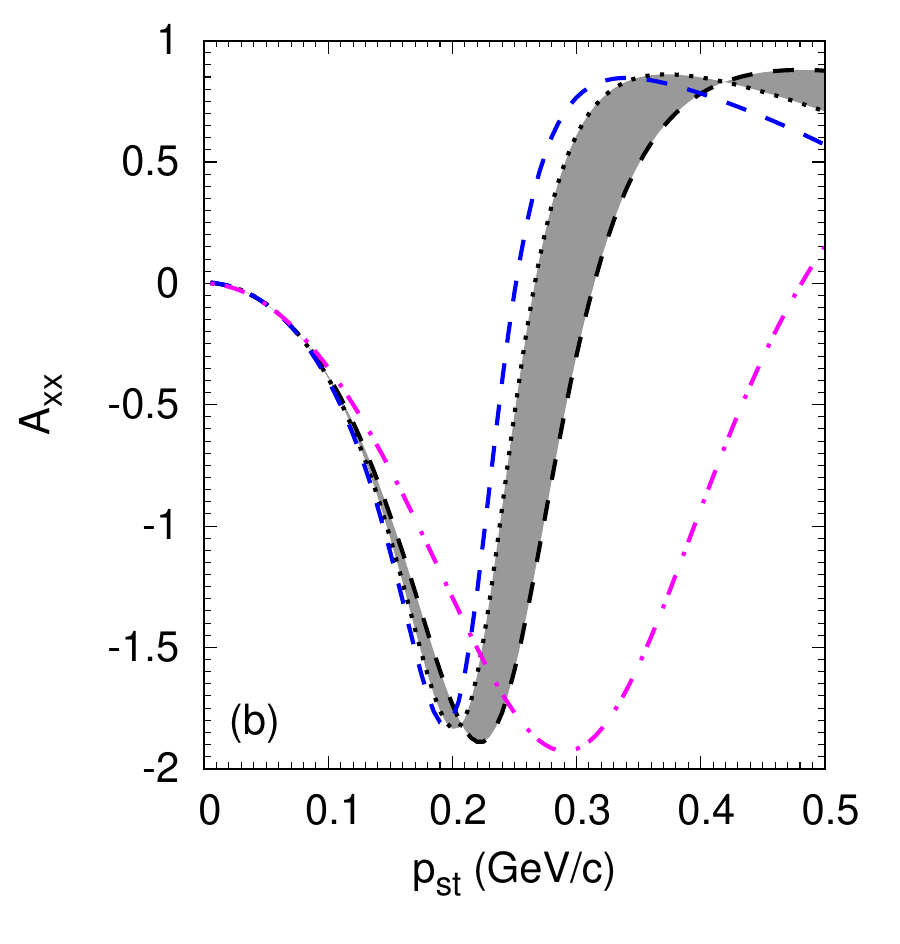} &
       \includegraphics[scale = 0.3]{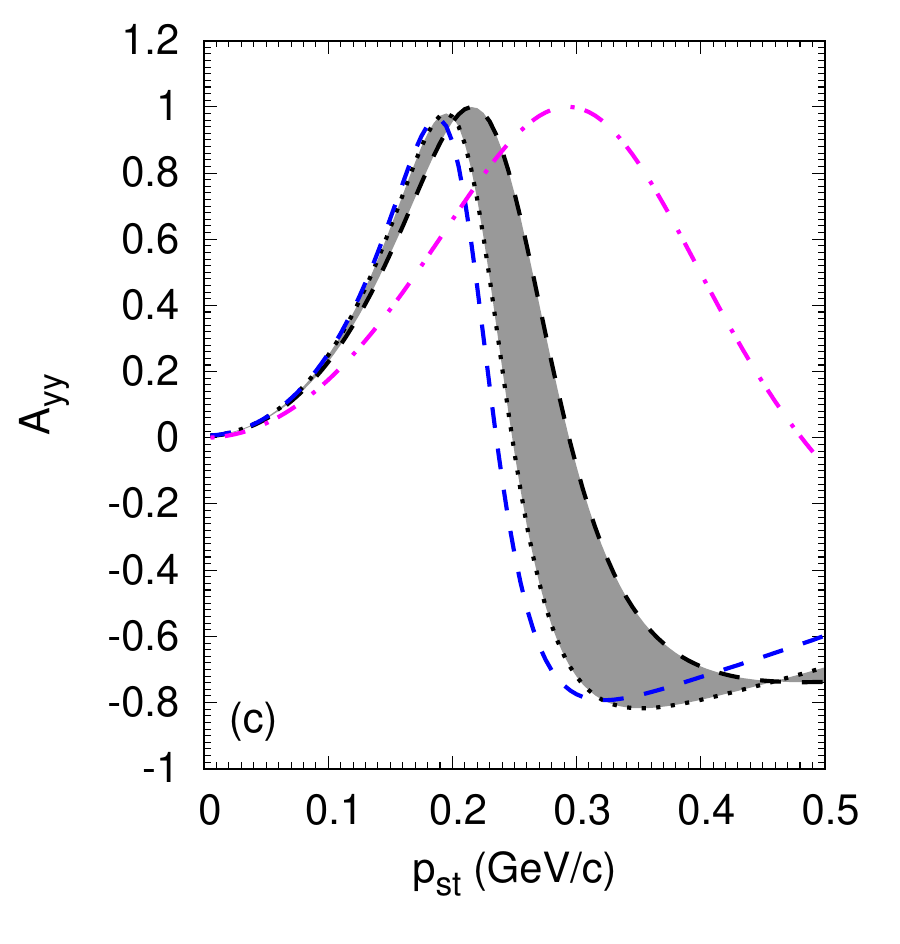} \\
       \includegraphics[scale = 0.3]{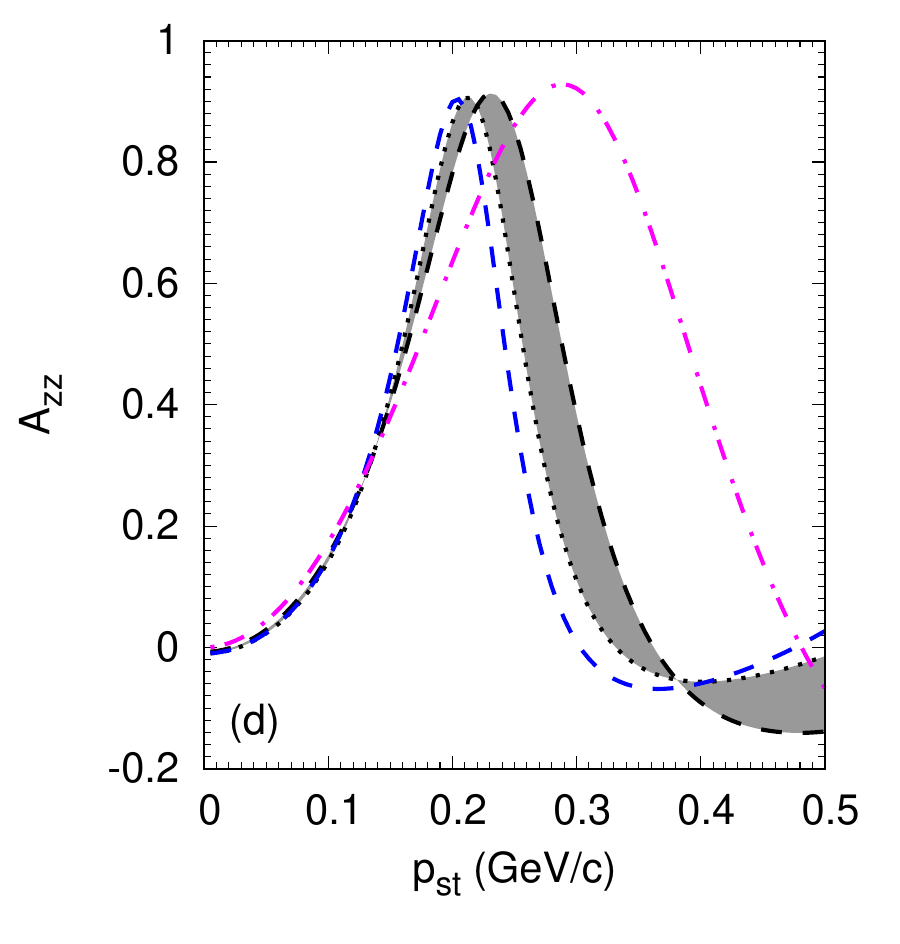} &
       \includegraphics[scale = 0.3]{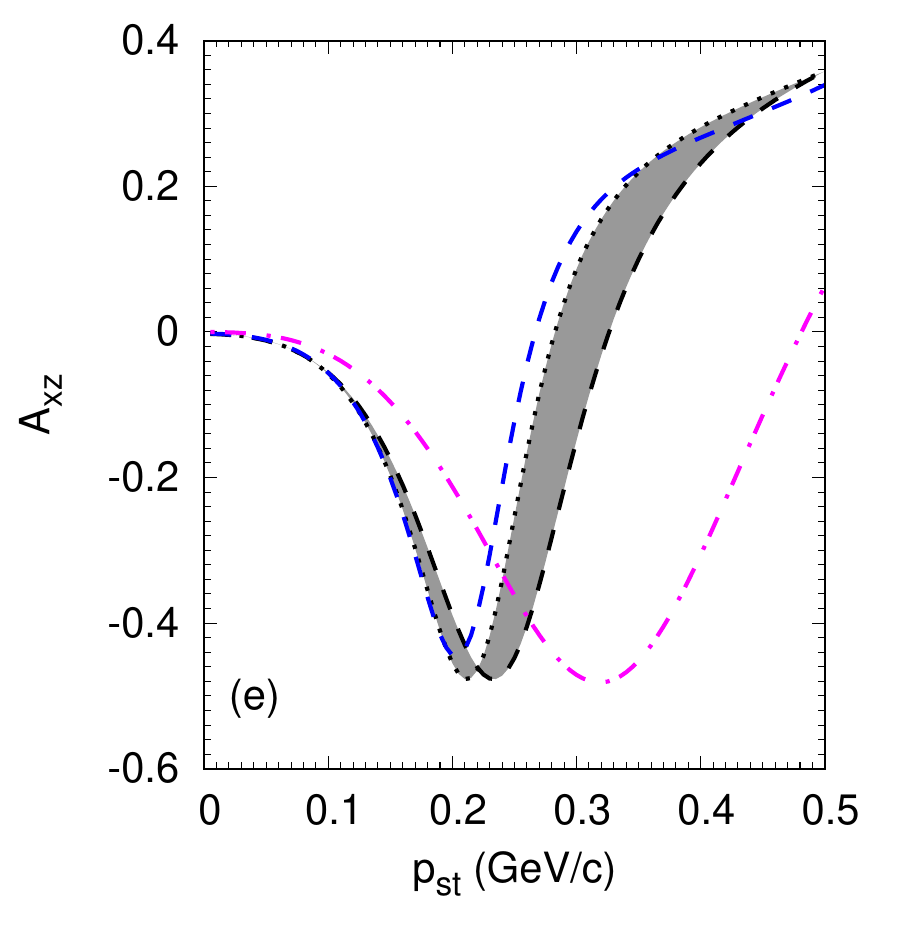} & \\
    \end{tabular}
  \end{center}
  \caption{\label{fig:pst-dep} DVAP and DTAP vs transverse momentum
    of spectator neutron for fixed relative azimuthal angle $\phi=180$ deg..
    Magenta dash-dotted line -- IA calculation.
    Blue dashed line -- GEA calculation.
    The gray band with borders given by the mass denominator of the coherence length,
    $\Delta M^2 = 0.7$ GeV$^2$ (black dashed line) and 3 GeV$^2$ (black dotted line)
    -- the calculation with CT effects.
  }
\end{figure}
Fig.~\ref{fig:pst-dep} shows the components $A_y, A_{xx}, A_{yy}, A_{zz}$ and $A_{xz}$ as functions of the transverse momentum of spectator neutron
at $\phi=180$ deg. Since all particles move in the $(x,z)$ plane, $A_x=A_z=A_{xy}=A_{yz}=0$, as follows from Eqs.(\ref{Avec}),(\ref{Atens}).
The component $A_y$ becomes finite when ISI/FSI are taken into account. At $p_{st}=0$ all components disappear which is the consequence
of the $D$-wave dominance. The extrema of the tensor components experience a shift from $p_{st} \simeq 0.3$ GeV/$c$
to $p_{st} \simeq 0.2$ GeV/$c$ due to ISI/FSI in the GEA calculations while the width of the $p_{st}$-dependence becomes smaller.
CT effects grow with $p_{st}$ and become most pronounced at $p_{st} \simeq 0.3$ GeV/$c$.

\begin{figure}[t!]
  \begin{center}
    \begin{tabular}{ccc}
       \includegraphics[scale = 0.3]{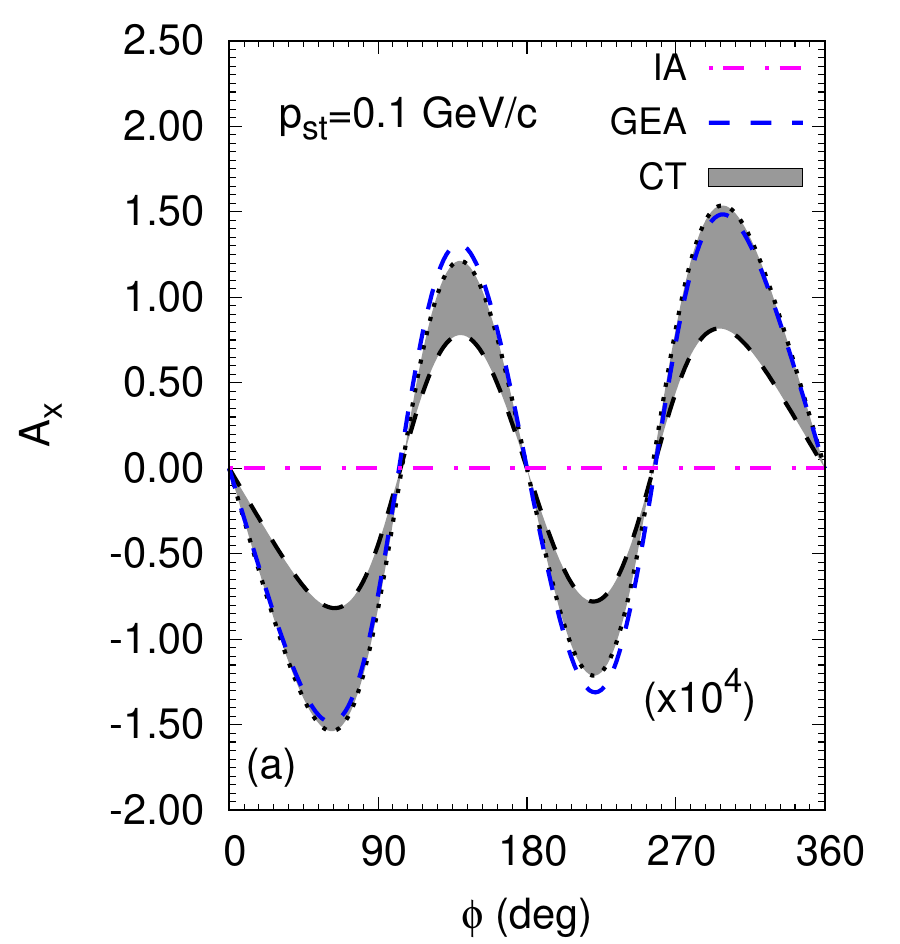} &
       \includegraphics[scale = 0.3]{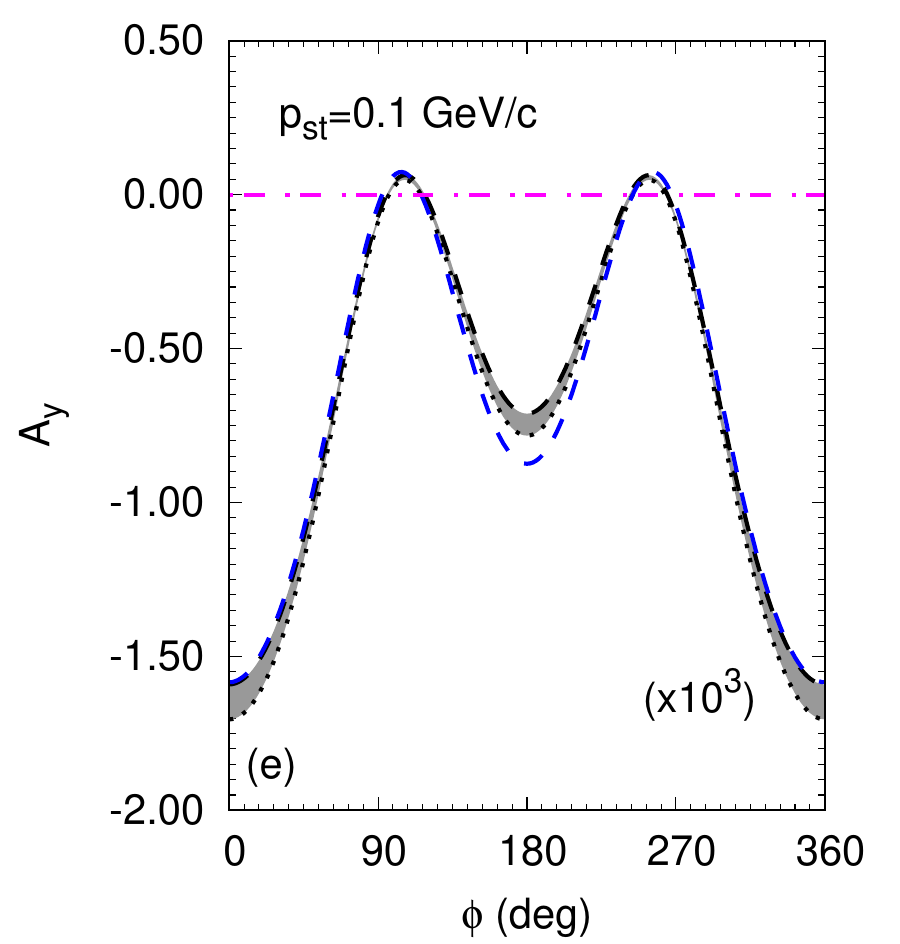} &
       \includegraphics[scale = 0.3]{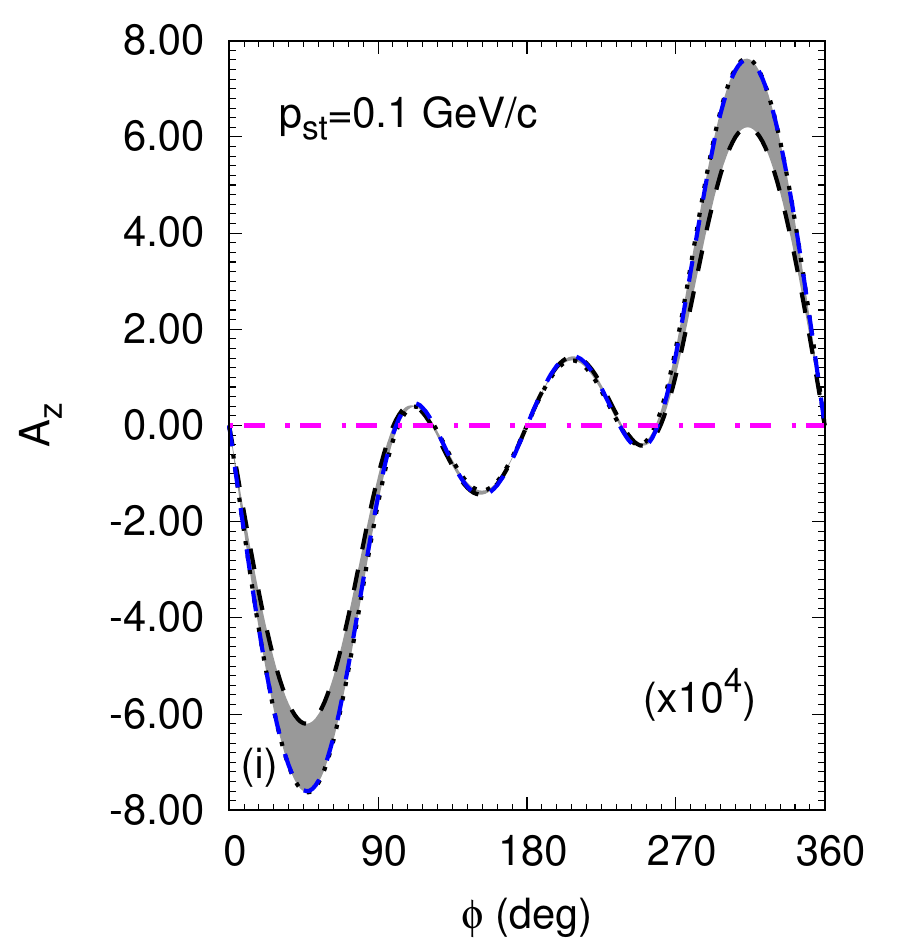} \\
       \includegraphics[scale = 0.3]{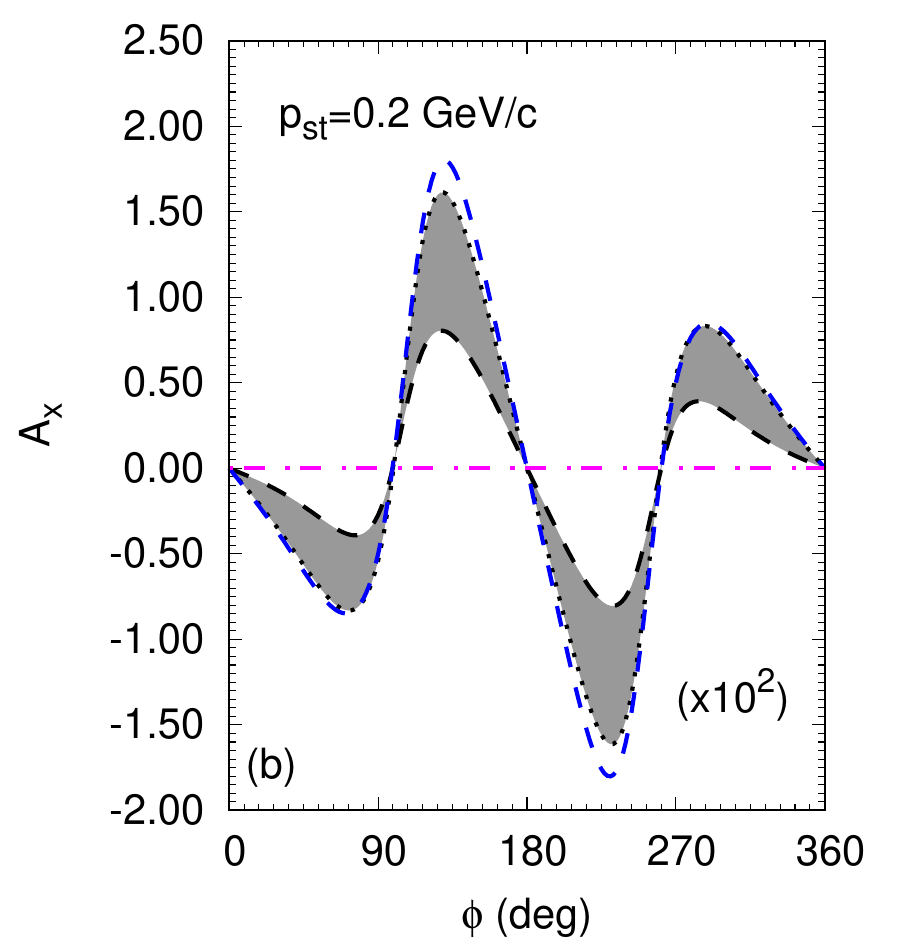} &
       \includegraphics[scale = 0.3]{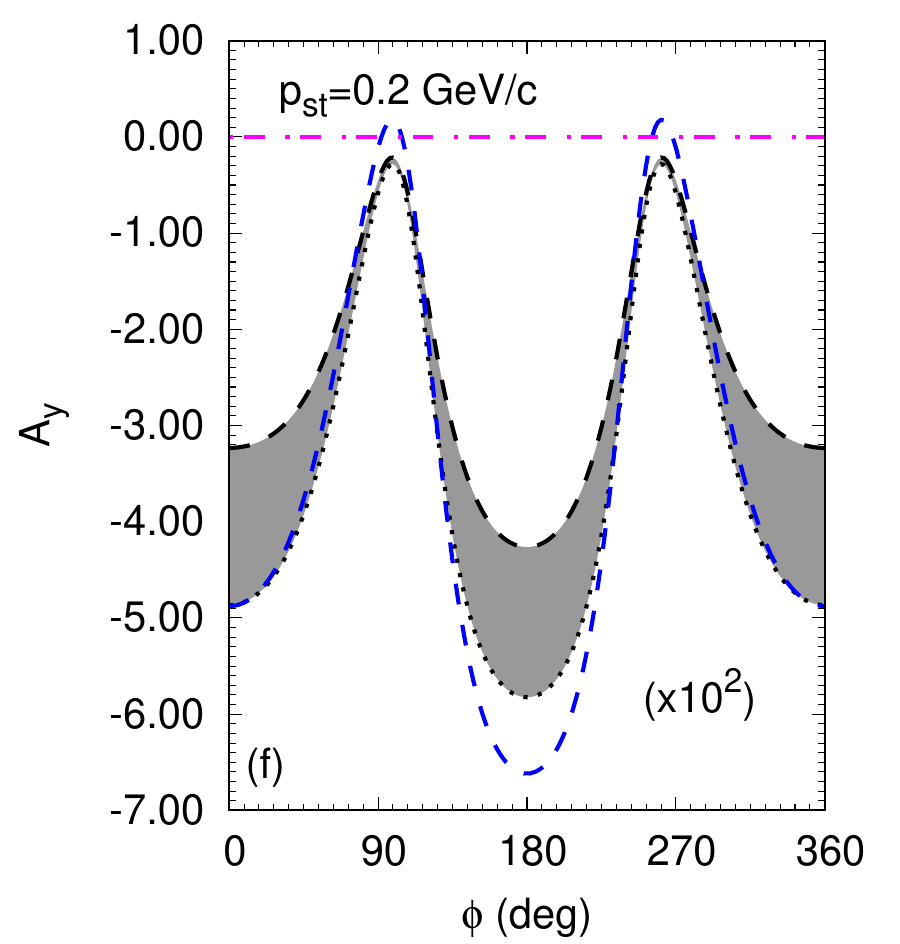} &
       \includegraphics[scale = 0.3]{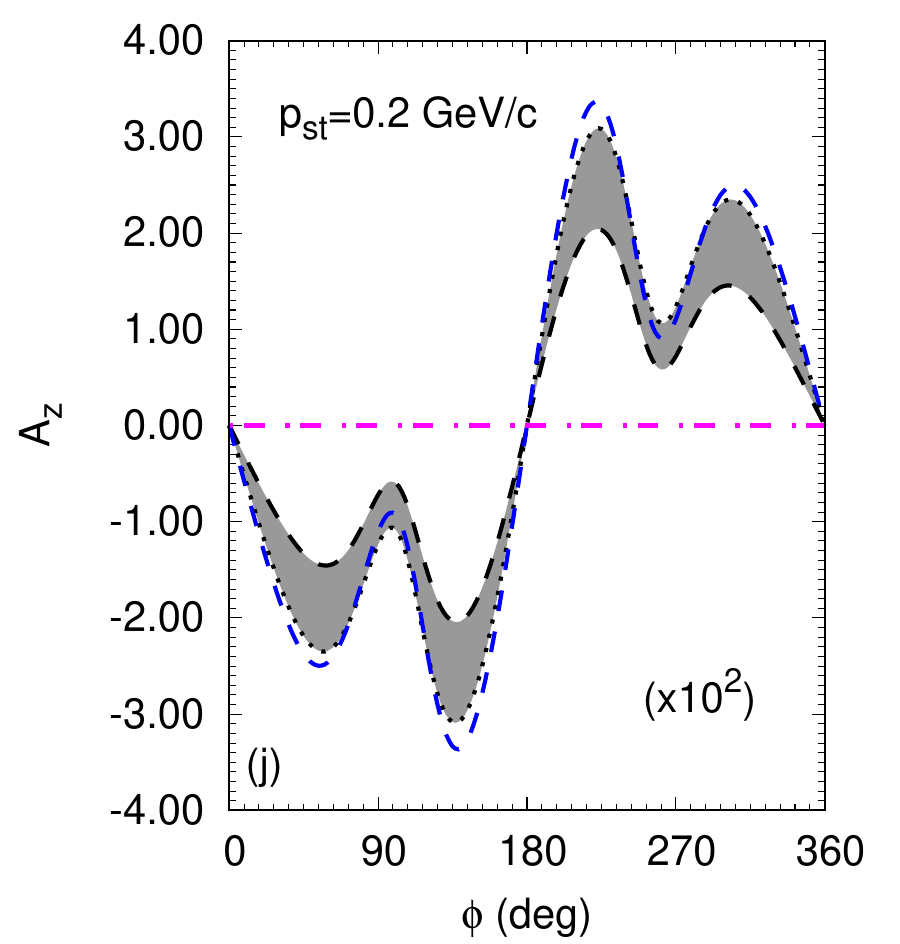} \\
       \includegraphics[scale = 0.3]{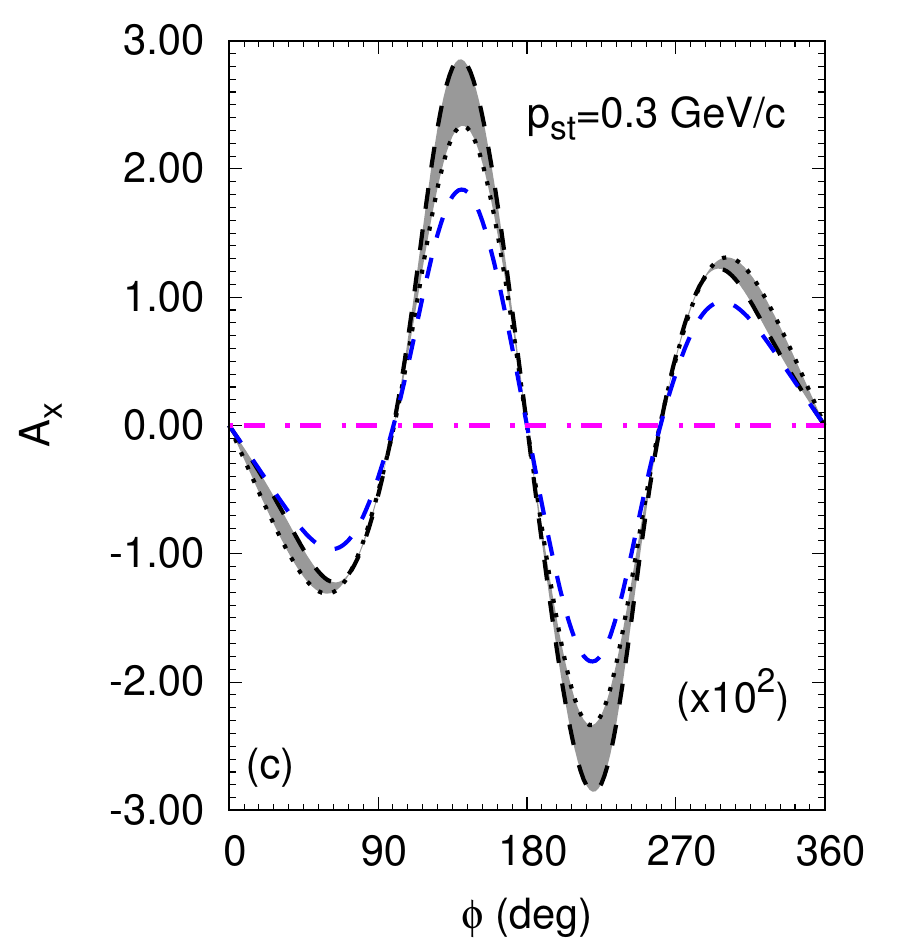} &
       \includegraphics[scale = 0.3]{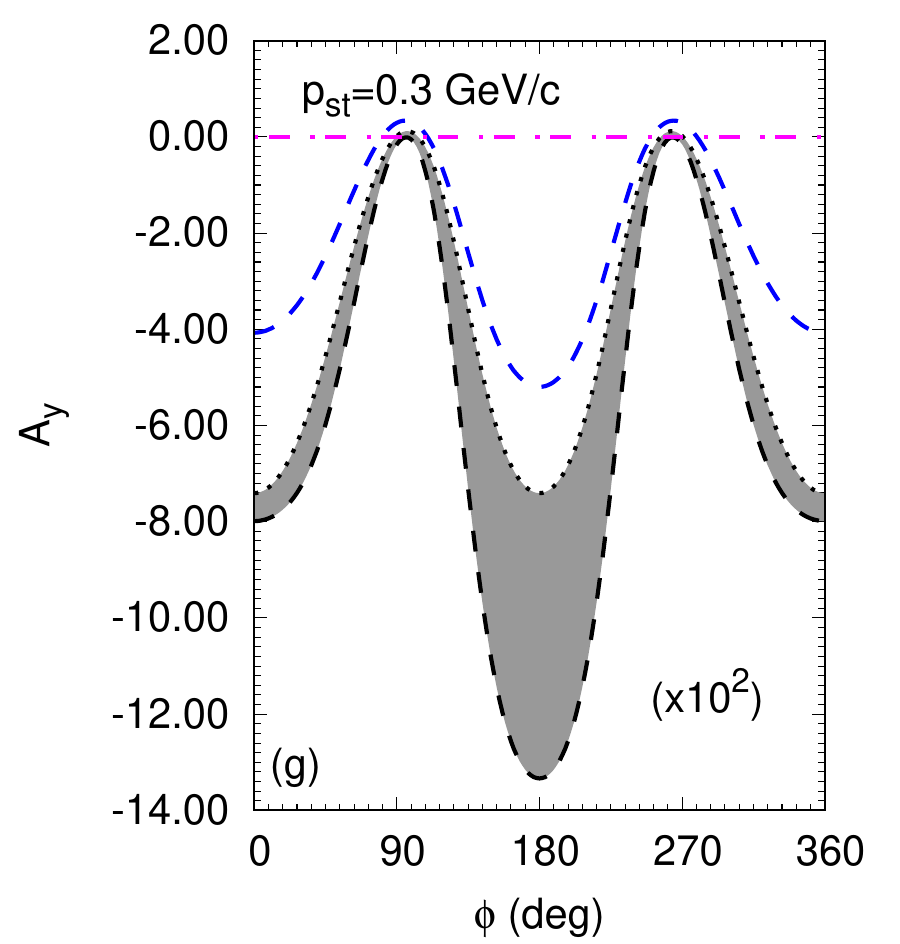} &
       \includegraphics[scale = 0.3]{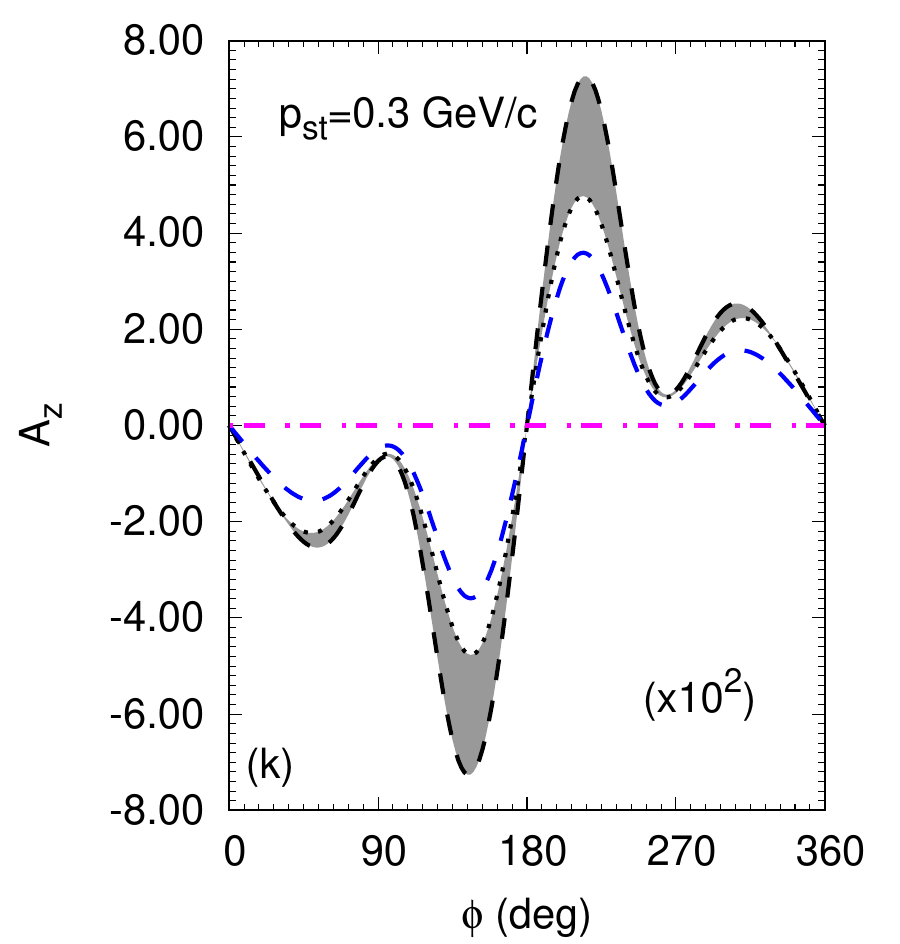} \\
       \includegraphics[scale = 0.3]{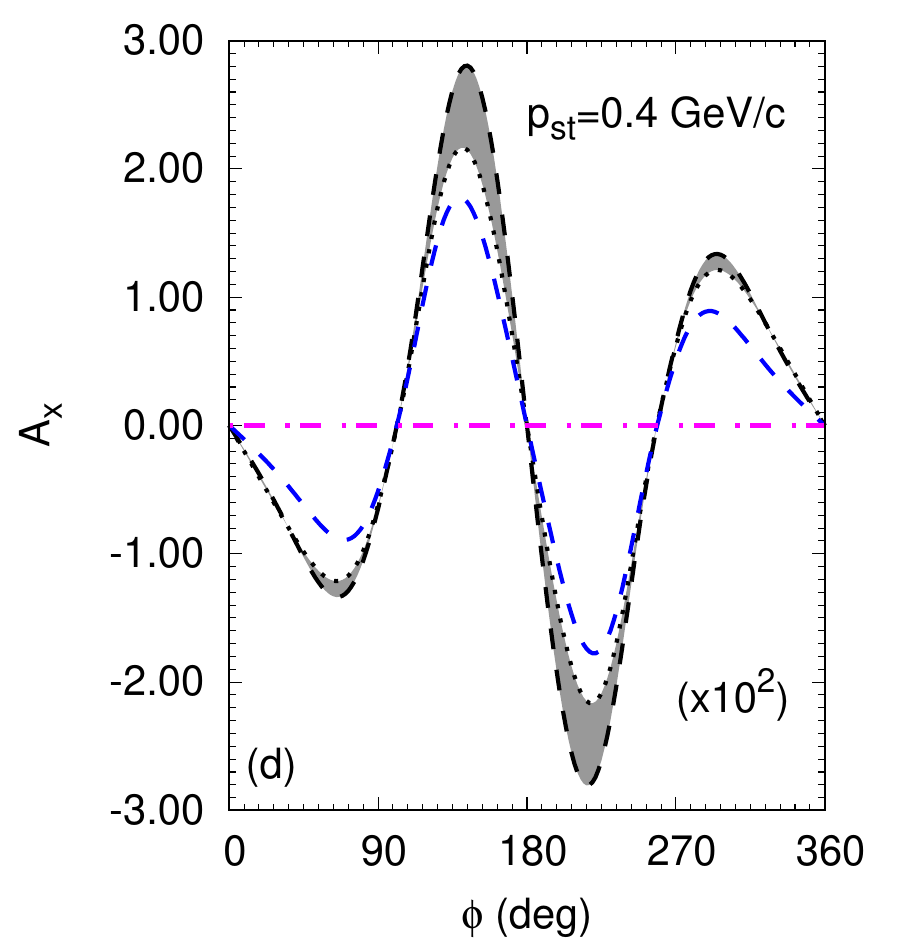} &
       \includegraphics[scale = 0.3]{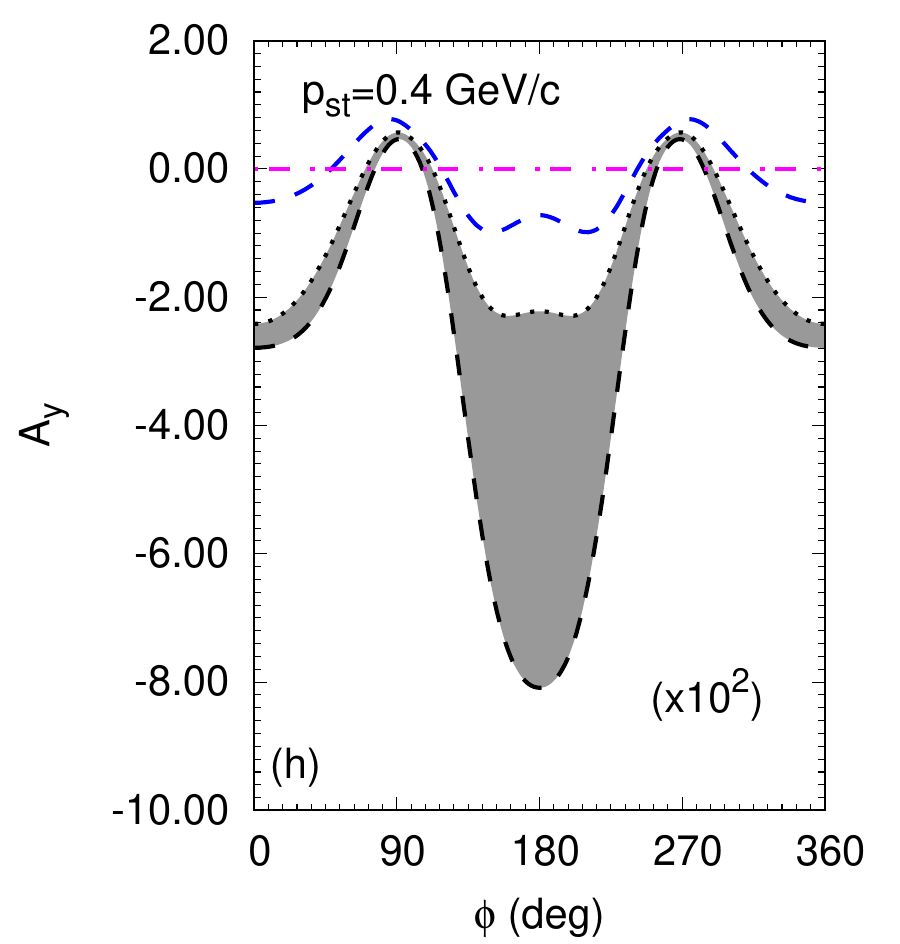} &
       \includegraphics[scale = 0.3]{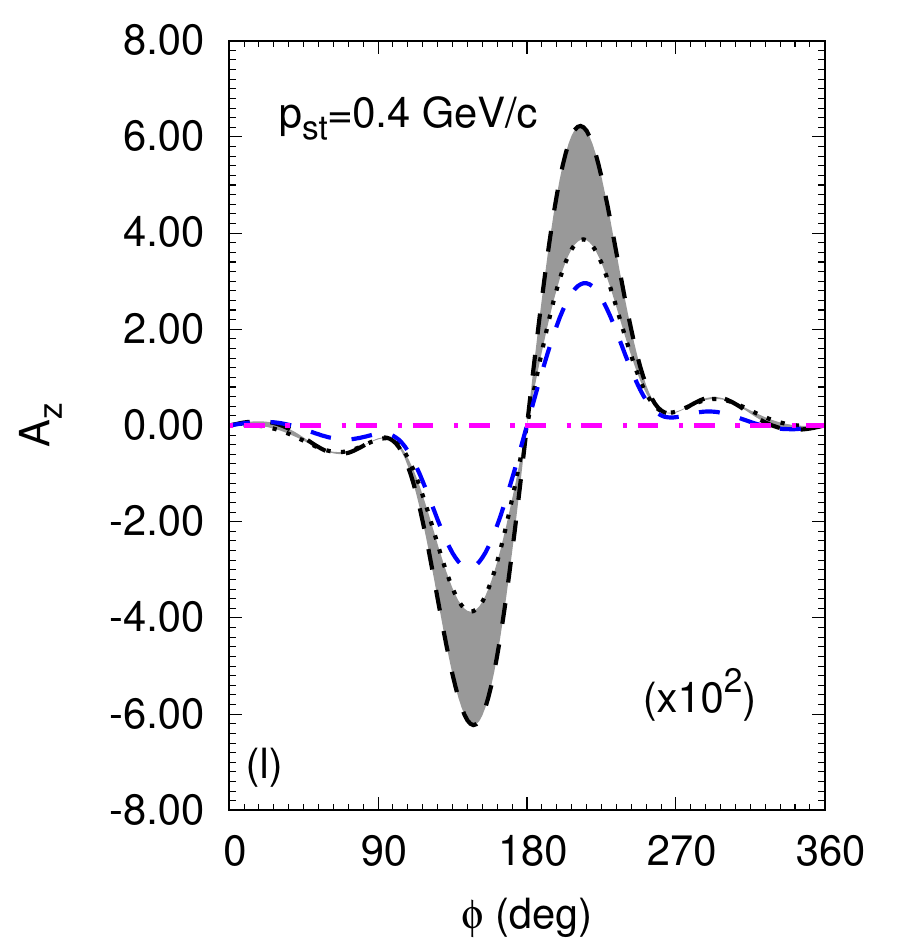} \\
    \end{tabular}
  \end{center}
  \caption{\label{fig:Av_phi-dep} DVAP as a function of relative azimuthal angle $\phi$
    between scattered proton and spectator neutron at different transverse momenta of spectator neutron.
    Plotted results are multiplied by the factors given in parentheses. 
    Line notation is the same as in Fig.~\ref{fig:pst-dep}.}
\end{figure}
Fig.~\ref{fig:Av_phi-dep} presents the azimuthal dependence of the DVAP. In line with Eq.(\ref{Avec}),
the components $A_x$ and $A_z$ are antisymmetric with respect to reflection in the $(x,z)$ plane while the component
$A_y$ is symmetric. The components of DVAP do not exceed 0.15 by absolute value. However, they exhibit quite significant
CT effects, in-particular, $A_y$ at large transverse momenta.

\begin{figure}[t!]
  \begin{center}
    \begin{tabular}{cccc}
       \includegraphics[scale = 0.25]{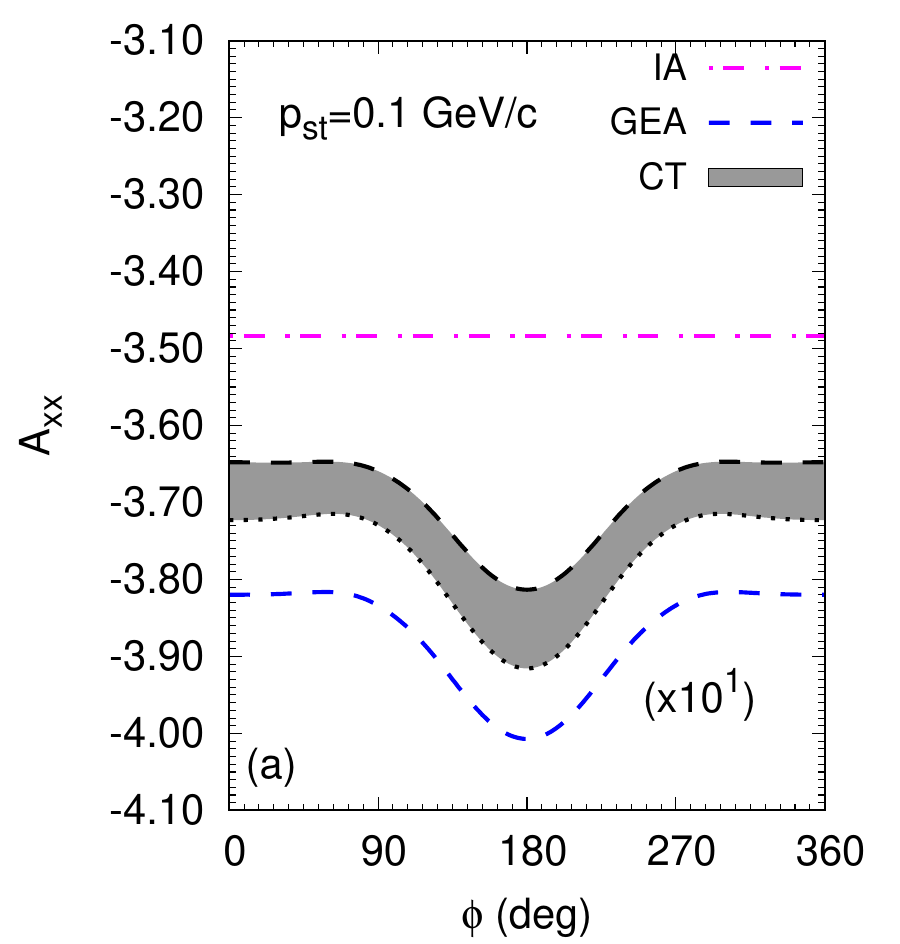} &
       \includegraphics[scale = 0.25]{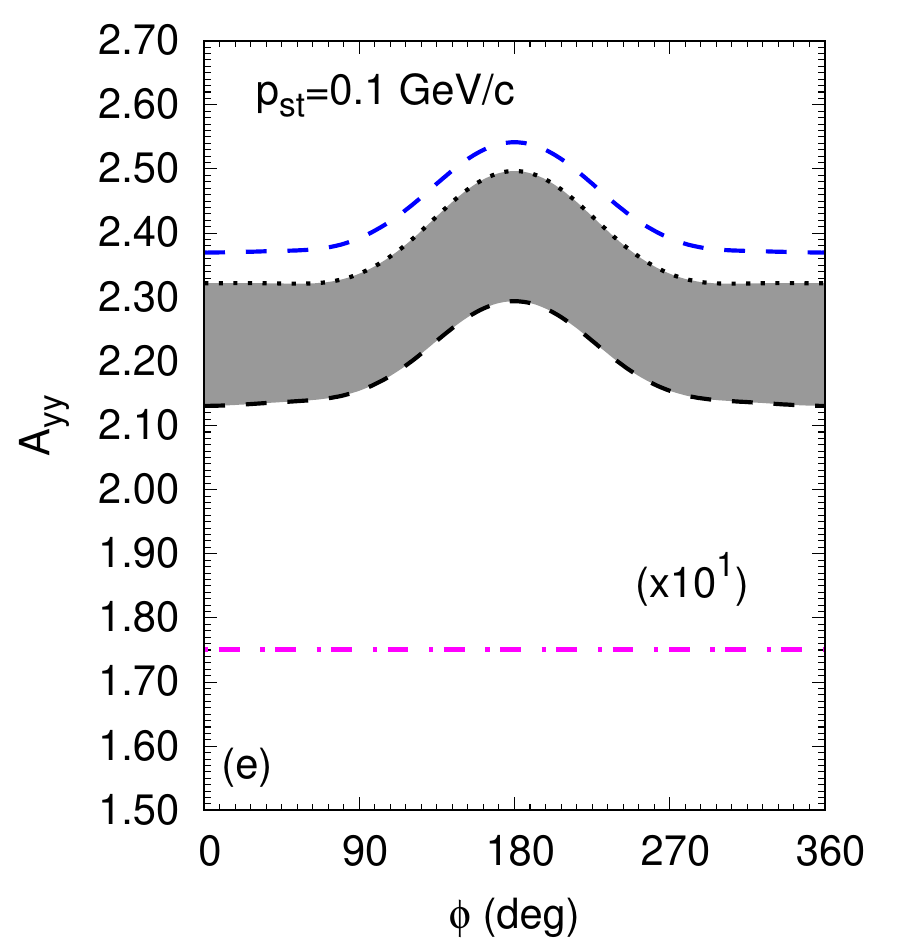} &
       \includegraphics[scale = 0.25]{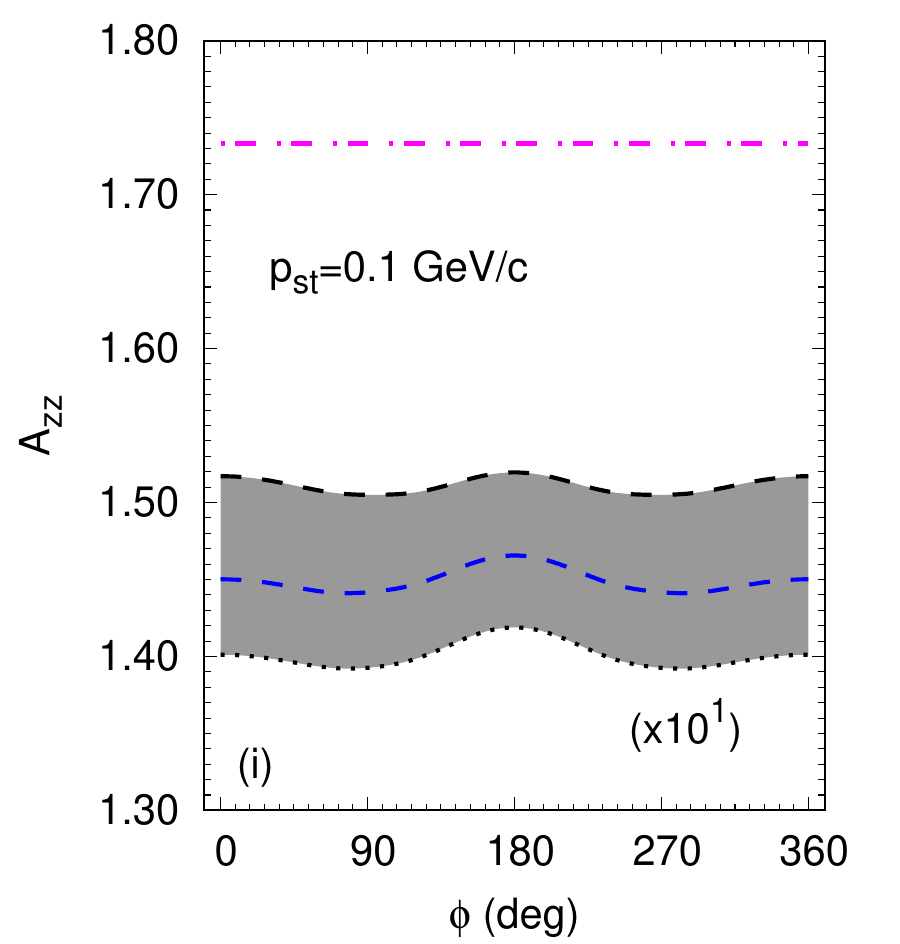} &
       \includegraphics[scale = 0.25]{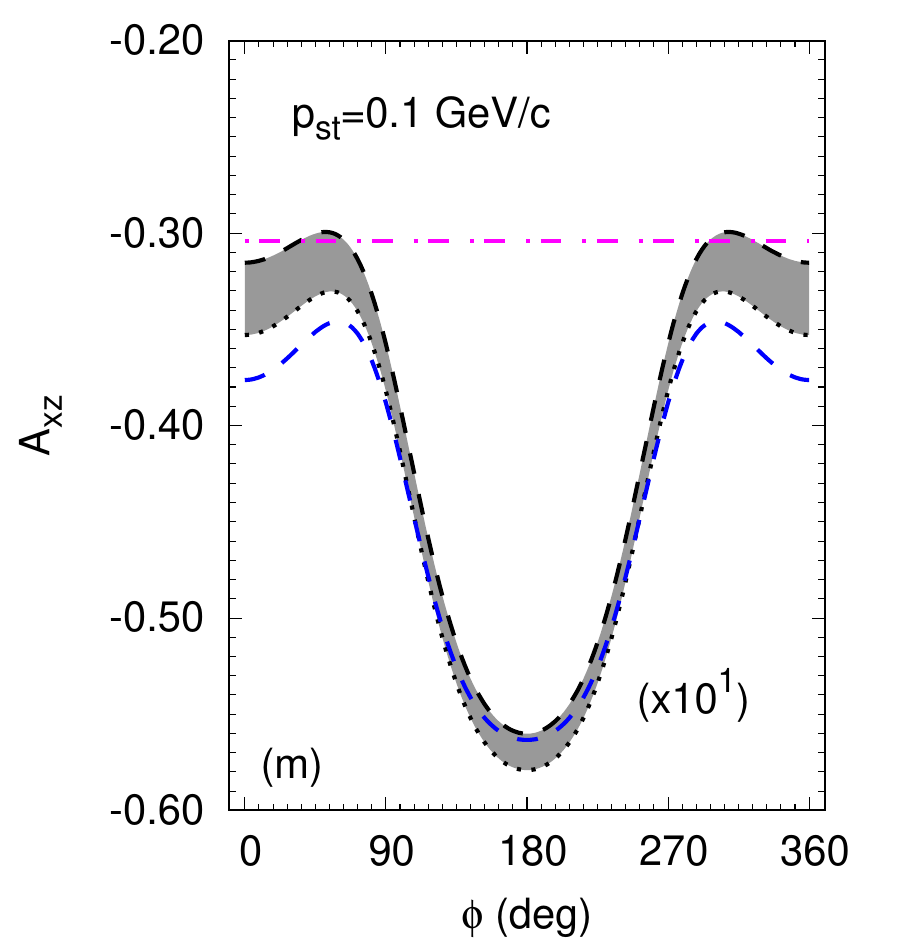} \\
       \includegraphics[scale = 0.25]{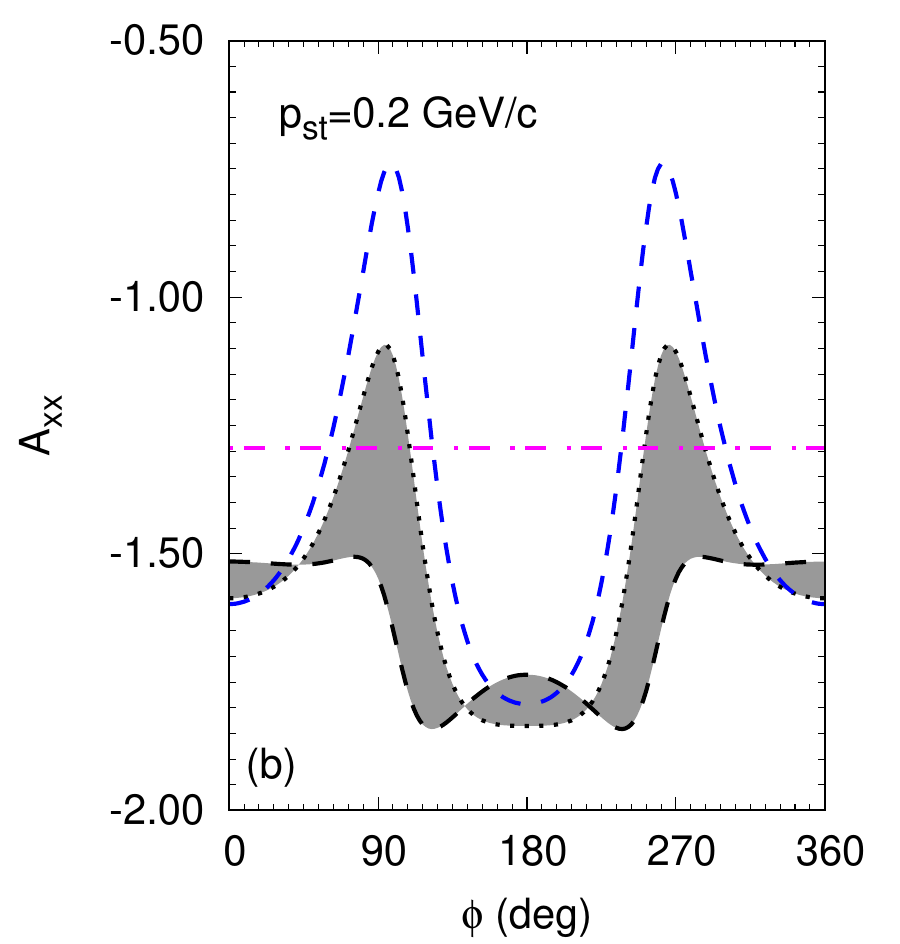} &
       \includegraphics[scale = 0.25]{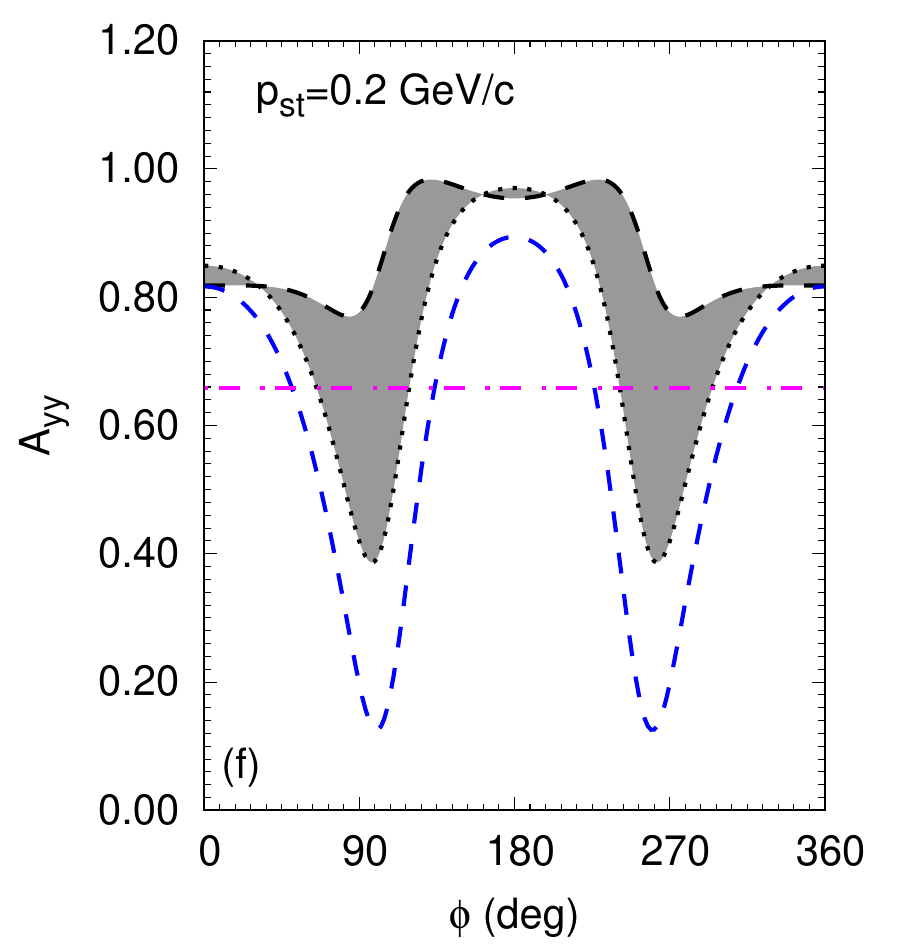} &
       \includegraphics[scale = 0.25]{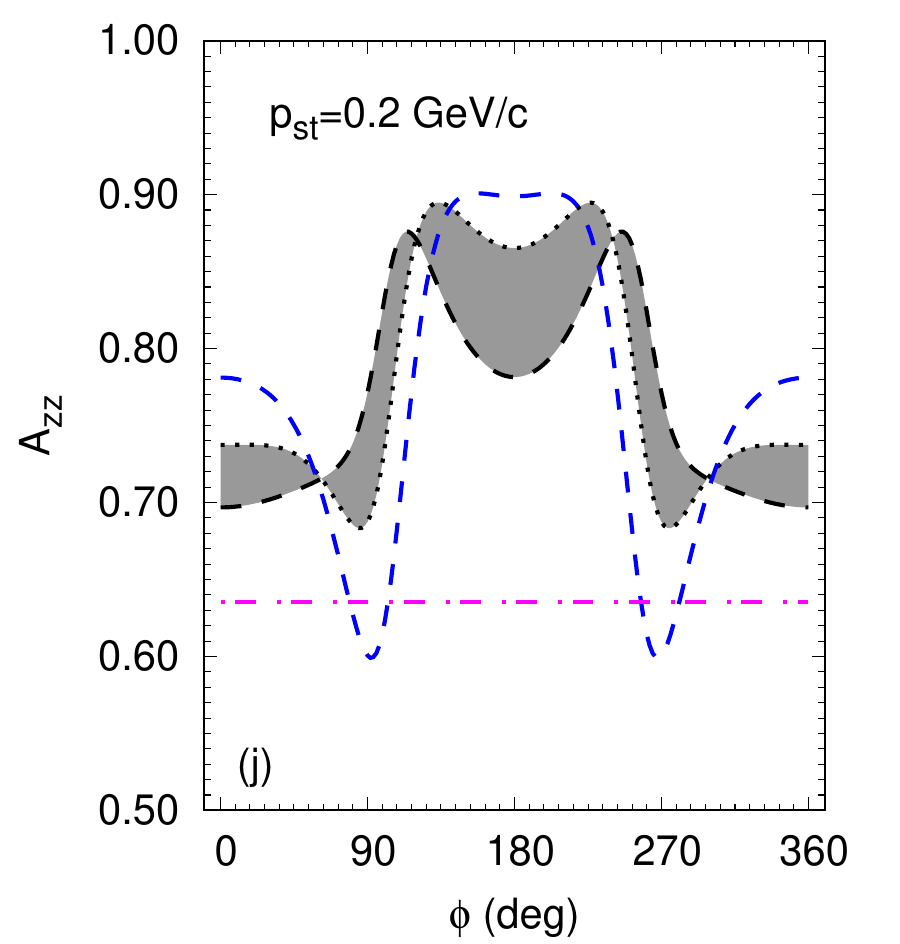} &
       \includegraphics[scale = 0.25]{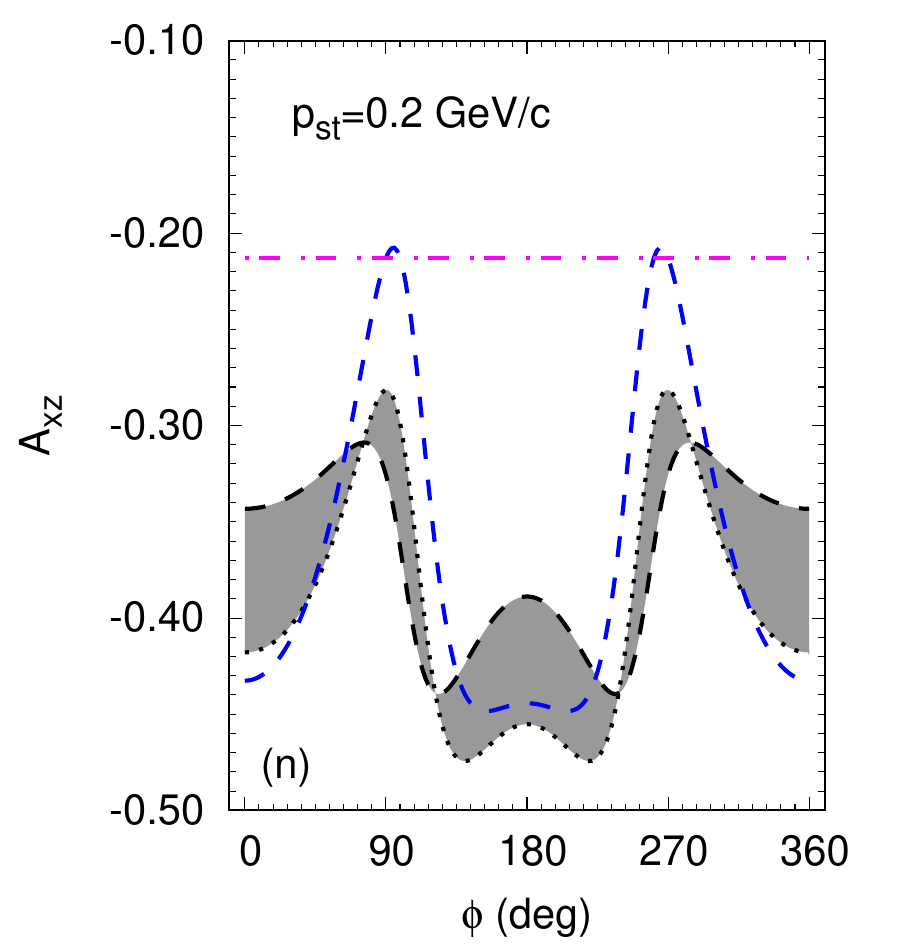} \\
       \includegraphics[scale = 0.25]{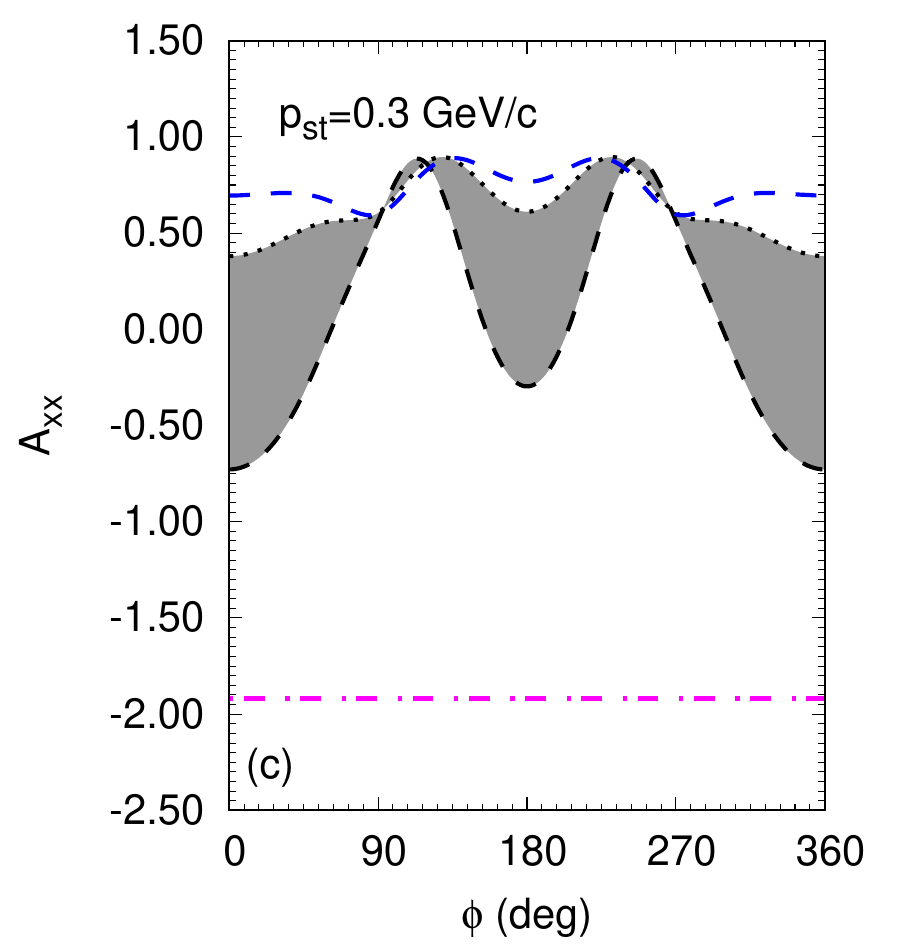} &
       \includegraphics[scale = 0.25]{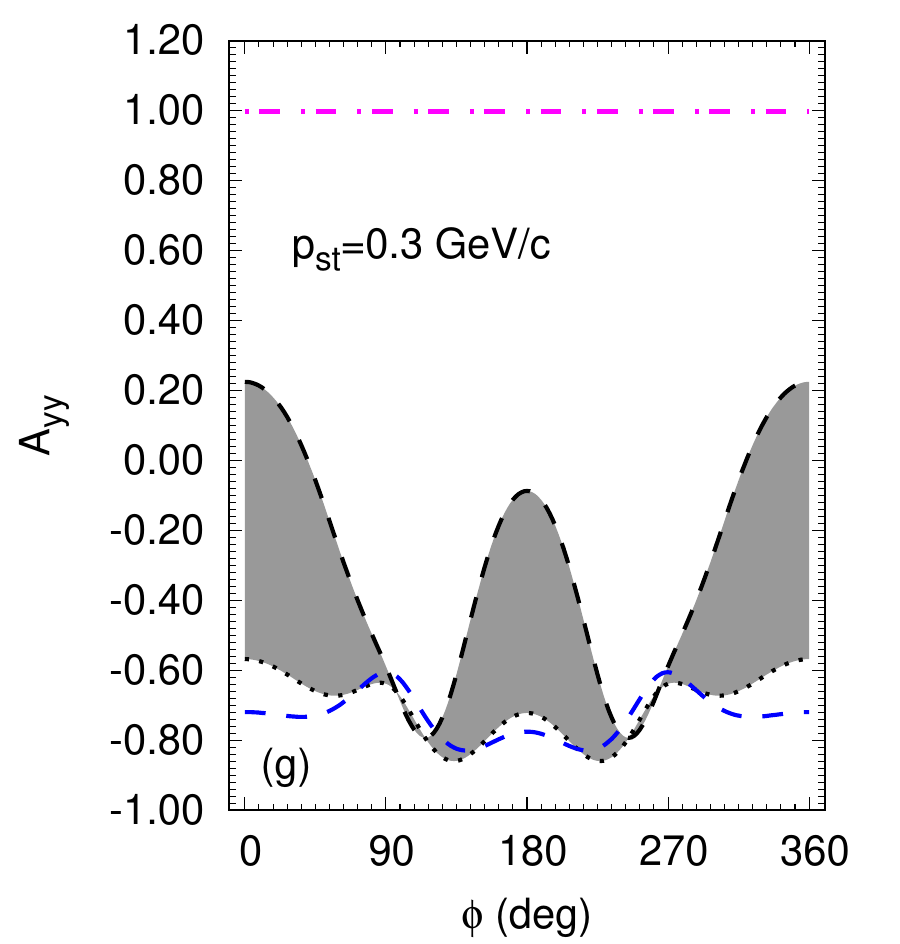} &
       \includegraphics[scale = 0.25]{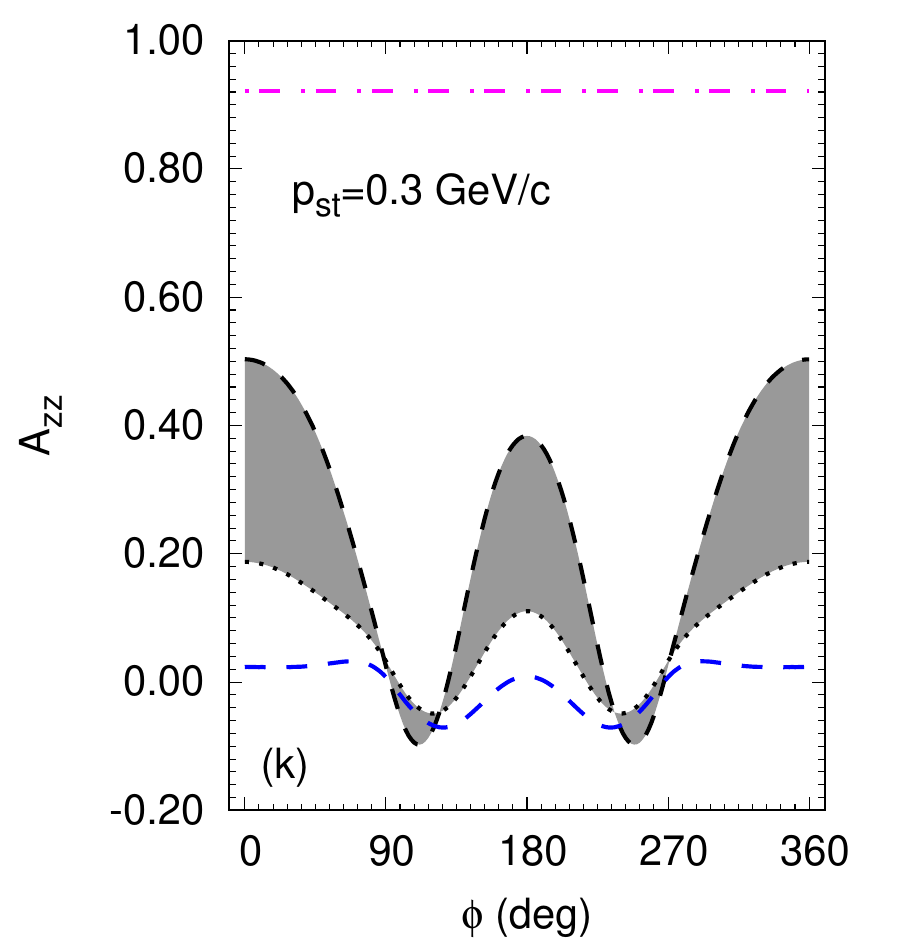} &
       \includegraphics[scale = 0.25]{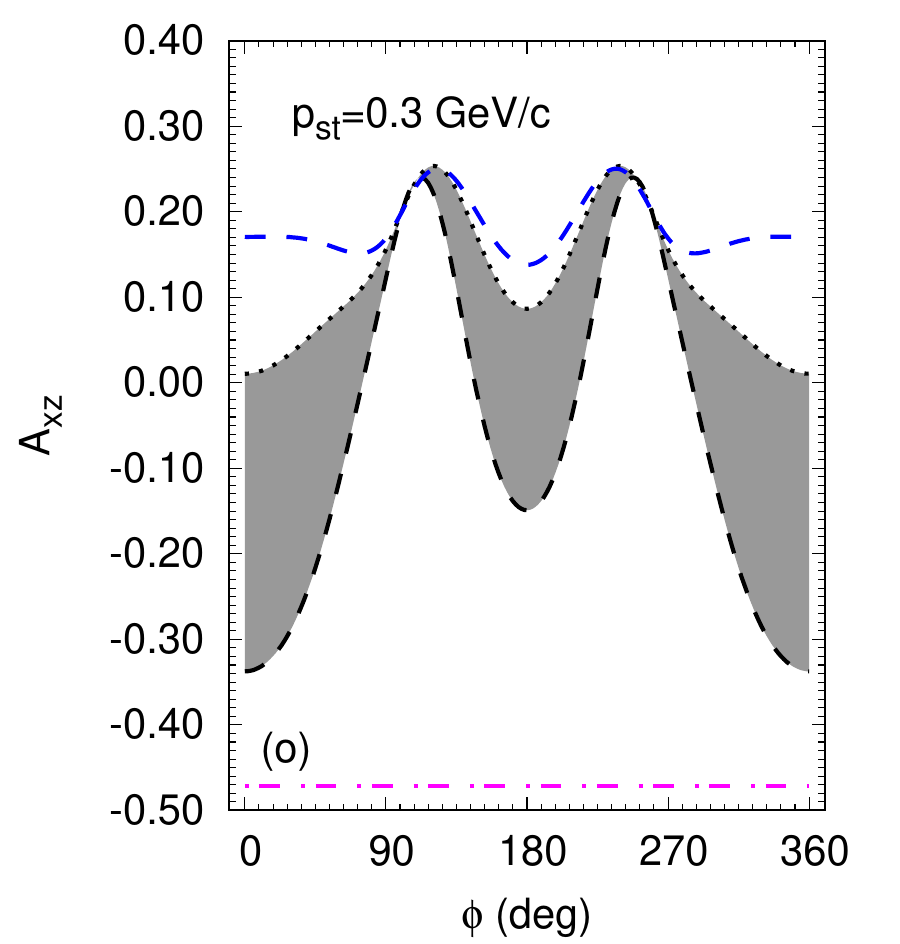} \\
       \includegraphics[scale = 0.25]{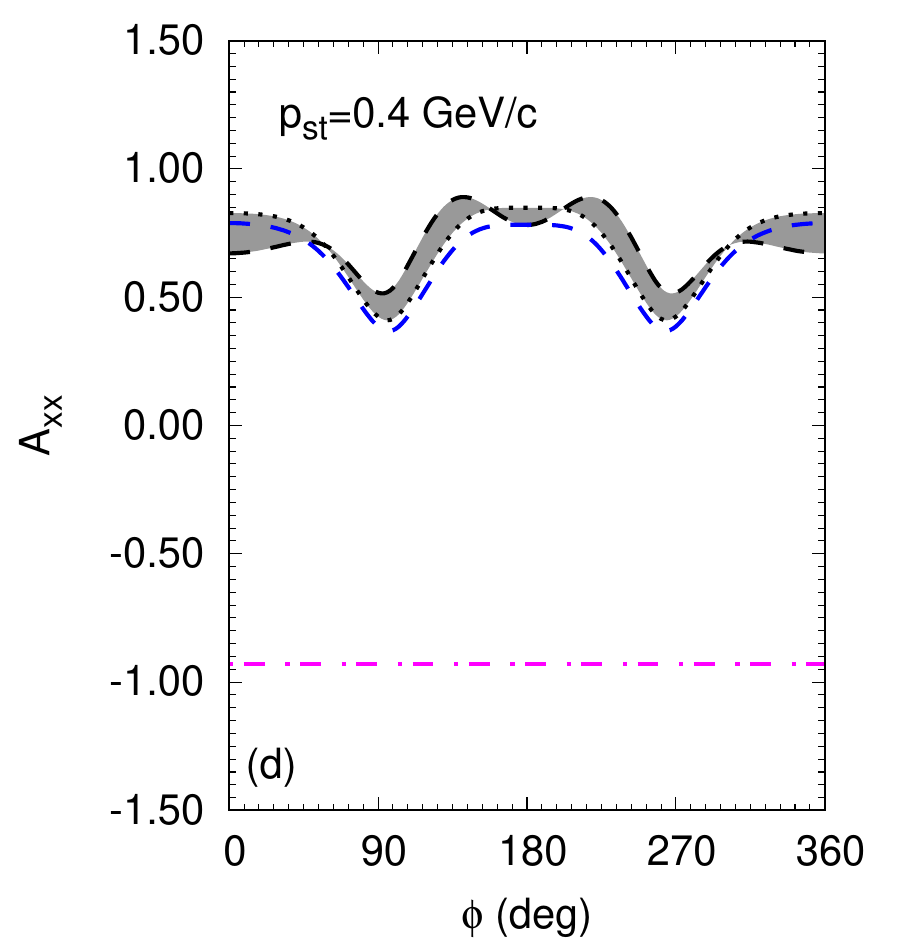} &
       \includegraphics[scale = 0.25]{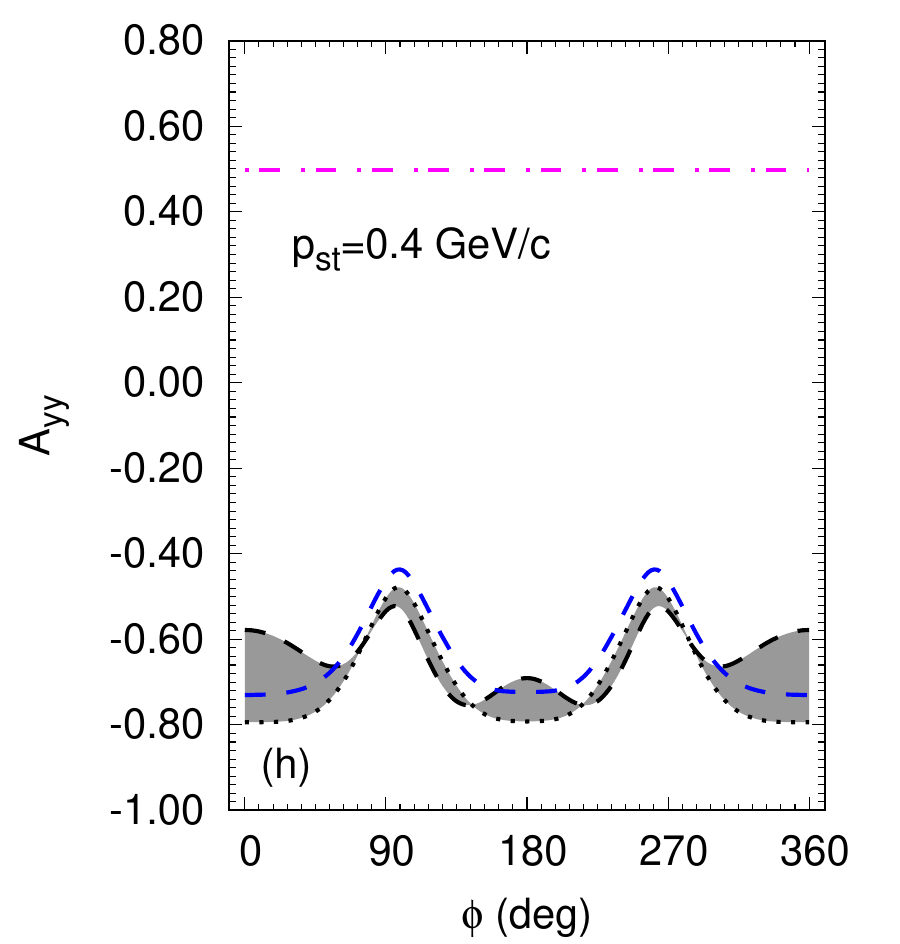} &
       \includegraphics[scale = 0.25]{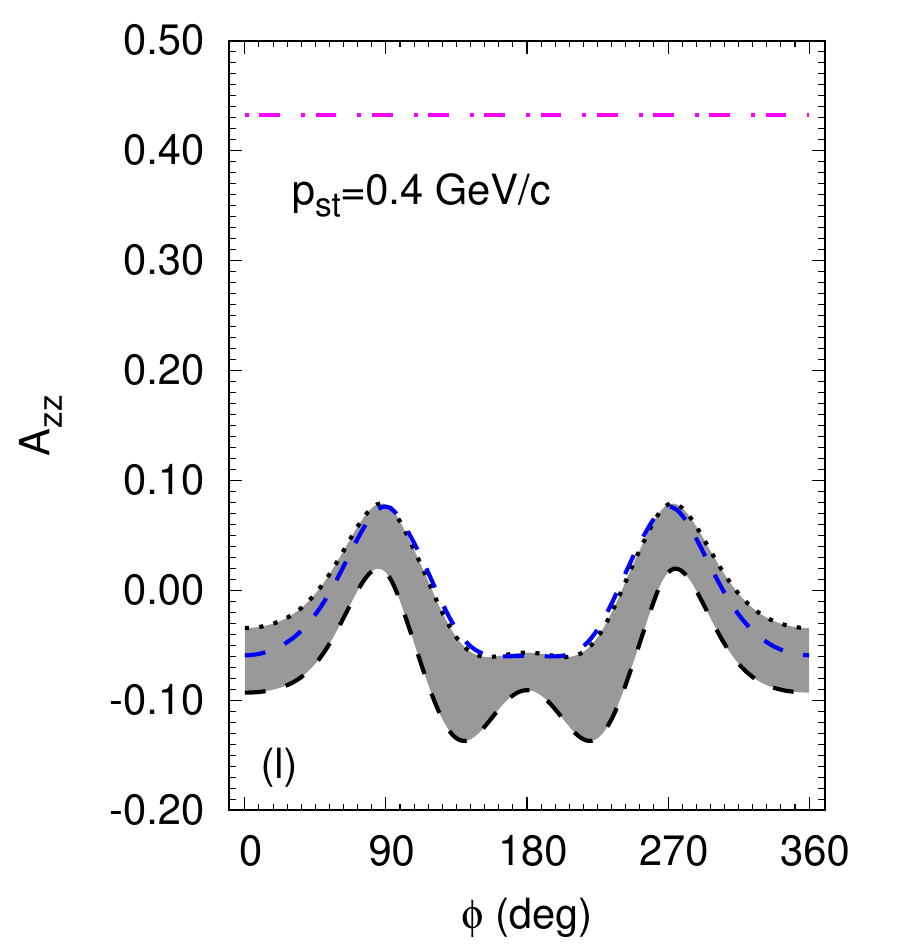} &
       \includegraphics[scale = 0.25]{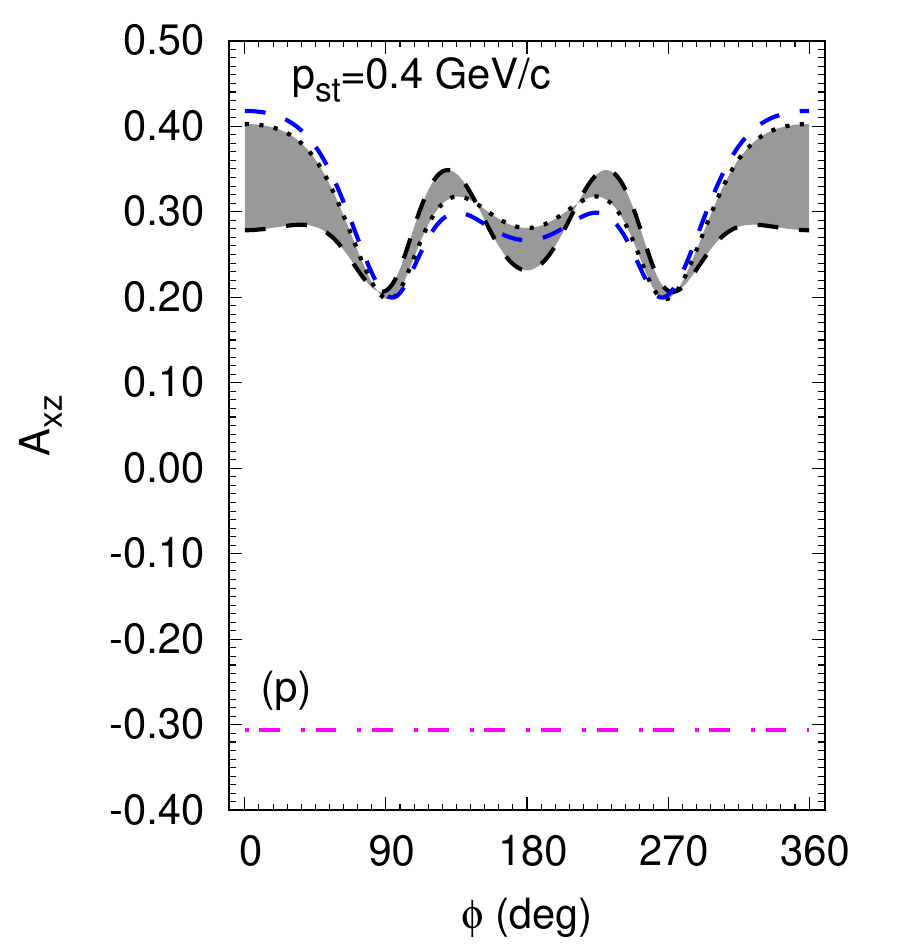} \\
    \end{tabular}
  \end{center}
  \caption{\label{fig:At_phi-dep} Same as Fig.~\ref{fig:Av_phi-dep}, but for the $A_{xx}, A_{yy}, A_{zz}$ and $A_{xz}$ components of DTAP.}
\end{figure}

The azimuthal dependence of the most significant components of DTAP is shown in Fig.~\ref{fig:At_phi-dep}.
In accord with Eq.(\ref{Atens}), the diagonal components as well as the $A_{xz}$ component are symmetric
with respect to reflection in the $(x,z)$ plane.
(The $A_{xy}$ and $A_{yz}$ components are antisymmetric with respect to reflection in the $(x,z)$ plane.
These components are much smaller, not exceeding $\sim 0.2$ by absolute value, and not shown.)
The azimuthal behavior of the components $A_{yy}$ and $A_{zz}$ is quite similar.
Since the spectator neutron moves almost along the positive $x$-direction, the $A_{xx}$ component is out of phase with $A_{yy}$ and $A_{zz}$ ones.
The components of DTAP show a pronounced variation with the angle $\phi$ due to ISI/FSI effects
which increase with increasing neutron transverse momentum up to $p_{st} \simeq 0.3$ GeV/$c$ and then saturate.               
CT strongly affects DTAP since the latter is governed by the $D$-wave component
of the deuteron wave function which selects small relative distance between proton and neutron in the deuteron, thus,
reducing the expansion of the point-like quark configurations. 
The influence of CT is strongest at $p_{st} \simeq 0.3$ GeV/$c$ for in-plane kinematics, i.e. for $\phi=0$ and $180$ deg.

\section{Summary}
\label{summary}

Calculations for the $d(p,2p)n$ hard deuteron breakup at $p_{\rm lab}=15$ GeV/$c$ ($\sqrt{s_{NN}}=5.5$ GeV)
were performed on the basis of the GEA that includes ISI/FSI effects in the form
of multi-step amplitudes with soft rescattering of the incoming and outgoing protons on the spectator neutron. 
The focus was on the deuteron polarization observables. 
The effects of CT were included within the quantum diffusion model taking into account 
the interference of the small- and large-size $qqq$ configurations.

The ISI/FSI effects lead to significant modification of the behavior of the DVAP and DTAP
as functions of the transverse momentum of spectator neutron and
relative azimuthal angle between the scattered proton and the spectator neutron.
These dependencies are sensitive to the CT effects.
Experimental determination of the DVAP and DTAP provides an excellent opportunity
for both GEA testing and CT identification in hard $pd$ collisions.
Such measurements can be performed on the first stage
of SPD NICA operation by using $dd$ collisions where one of the colliding deuterons is non-polarized and serves as a source of the quasi-free proton
that experiences a hard collision with another polarized deuteron, see Refs.~\cite{Larionov:2025equ,Uzikov:2025uau,Datta:2026lsf}.
Furthermore, the ISI/FSI and possibly the onset of CT could be studied at lower energies at the JINR nuclotron
using a deuteron beam interacting with a proton target.

\bibliography{protvCT}

\end{document}